\documentclass[aps,pre,twocolumn,floatfix]{revtex4}

\usepackage[centertags]{amsmath}
\usepackage{amsfonts}
\usepackage{amssymb}
\usepackage{amsthm}
\usepackage{newlfont}
\usepackage{bbm}

\theoremstyle{plain}

\theoremstyle{definition}

\theoremstyle{remark}

\numberwithin{equation}{section}

\newcommand{\bbE}{{\mathbb E}}

\newcommand{\bbZ}{{\mathbb Z}}

\newcommand{\opunit}{\text{1}\kern-0.22em\text{l}}

\usepackage{graphicx}
\usepackage{bm}
\bibstyle{apsrev.bib}

\newcommand{\beqa}{\begin{eqnarray}}
\newcommand{\eeqa}{\end{eqnarray}}
\newcommand{\beq}{\begin{equation}}
\newcommand{\eeq}{\end{equation}}
\newcommand{\ave}[1]{\langle #1\rangle}
\newcommand{\ddd}{\partial}
\newcommand{\grad}{\vec{\nabla}}
\newcommand{\dive}{\vec{\nabla}\cdot}
\newcommand{\curl}{\vec{\nabla}\times}

\begin{document}
\setlength{\voffset}{6mm}

\title{On the enstrophy dissipation in two-dimensional turbulence}

\author{Marco Baiesi}
\author{Christian Maes}\email{christian.maes@fys.kuleuven.be}
\address{Instituut voor Theoretische Fysica, K.U.Leuven,
Celestijnenlaan 200D, B-3001, Belgium}
\date{\today}


\begin{abstract}
Insight into the problem of two-dimensional turbulence can be
obtained by an analogy with a heat conduction network. It allows
the identification of an entropy function associated to the
enstrophy dissipation and that fluctuates around a positive (mean)
value. While the corresponding enstrophy network is highly
nonlocal, the direction of the enstrophy current follows from the Second
Law of Thermodynamics. An essential parameter is the ratio $T_k\equiv
\gamma_k/(\nu k^2)$ of the intensity of driving $\gamma_k>0$ as a
function of wavenumber $k$, to the dissipation strength $\nu k^2$, where
$\nu$ is the viscosity. The enstrophy current flows from higher to
lower values of $T_k$, similar to a heat current from higher to
lower temperature.   Our probabilistic analysis of the enstrophy
dissipation and the analogy with heat conduction thus complements
and visualizes the more traditional spectral arguments for the
direct enstrophy cascade.
We also show a fluctuation symmetry in the distribution of the
total entropy production which relates the probabilities of direct
and inverse enstrophy cascades.
\end{abstract}

\maketitle

\section{Introduction}
\label{sec:intro}

Three-dimensional turbulence displays an inertial range, in which
energy is transferred from the spatial scales at which it is
introduced into the system down to small scales, where it is
finally dissipated by viscous forces. The standard picture of
turbulence in two dimensions  is qualitatively different.
Following the pioneering works of
Kraichnan~\cite{kraichnan67:_t2d,kraichnan71},
Leith~\cite{leith68:_t2d} and Batchelor~\cite{batchelor69:_t2d}
(KLB) on the two-dimensional Navier-Stokes equation for the
fluid velocity field, one expects an inverse energy cascade
from the forcing scales to large scales and simultaneously a
direct enstrophy cascade from the forcing scales to small scales.
The enstrophy is the variance of the vorticity, namely the
ensemble average of the squared curl of the velocity.

Two-dimensional turbulence has been a very active area of
theoretical, numerical and experimental
investigation~\cite{Tabeling02:_2d-turb,KellayGoldburg2002},
not only as an easier test case but also relevant to certain real
quasi-two-dimensional situations.
Examples include oceanic currents and atmospheric and geophysical
flows~\cite{Dritschel93:_atmos},
but two-dimensional
flow is also realized in laboratory
situations~\cite{Tabeling02:_2d-turb,KellayGoldburg2002}.
However, the picture of
two-dimensional turbulence remains not fully understood
and in fact, there are some limitations to the above classical KLB
scenario. For example, the standard enstrophy cascade disappears when
considering a bounded domain where only a monoscale forcing is
applied~\cite{tran3}.
Moreover, the mechanism of the direct enstrophy cascade and the
determining factor for the direction of the enstrophy current has
not been fully understood as a consequence of a more general
principle.  There have been recent clarifications, going
into the details of the physical mechanism, e.g.~\cite{eyink96,
chen03:_enstr_casc}, but
it seems interesting and natural to connect the situation also
with better understood scenario's and to be able to see the
enstrophy dissipation as the result of a more generally valid
principle.

In the present paper, we address the issue of the
enstrophy current and of its direction.  A very
close analogy with a two-dimensional heat conduction problem
provides new ingredients to understand the enstrophy
cascade in its full qualitative behavior.  It turns out, as will
be shown later, that the stochastically driven Navier-Stokes
equation for the vorticity can be mapped to a problem of heat
conduction: at each wavenumber $k$ a thermal reservoir is attached
with temperature $T_k=\gamma_k/(\nu k^2)$ where $\gamma_k$ is the
forcing strength and $\nu$ is the viscosity.  From the Second Law
of Thermodynamics, that will be derived in its detailed version,
follows that enstrophy is dissipated as heat flows: from higher to
lower temperature, or here, when $\gamma_k$ is peaked around some
small mode $k$, from small to large wavenumbers.  In other words,
the origin and the direction of the enstrophy flux is simply
and directly a consequence of the Second Law applied to the
enstrophy.  We also go beyond the study of the average enstrophy current
and discuss a symmetry in its fluctuations.
That estimates the probability of going backwards, i.e., the probability
 of an inverse enstrophy cascade.
At the same time, we obtain for the first time a steady state
fluctuation theorem in the context of turbulence.

In the next section, we start by reminding the reader of the
standard picture of two-dimensional turbulence.
In Section~\ref{sec:analogy} comes the analogy with heat conduction.
From it follows the final analysis of the enstrophy dissipation in
Section~\ref{sec:dissip}.
The main general consequences and conclusions are taken in
Section~\ref{sec:concl}.

The paper will describe the arguments and analogues in a formal
way, avoiding however a fully rigorous mathematical analysis. The
main goal is indeed to point out a useful picture and analogy
which is sufficiently powerful to specify the enstrophy cascade.
To add the mathematical details and hypotheses is not believed to
be extremely difficult but only few remarks are added to guide the
mathematically inclined.

\section{Navier-Stokes equation}
\label{sec:NS}

The Navier-Stokes (NS) equation~\cite{lesieur87:book,frisch:book}
for the velocity field $\vec u(t,r)$ is \beq \frac{\ddd \vec
u}{\ddd t} + (\vec u \cdot  \grad) \vec u = \nu\triangle\vec u
-\grad p + \vec f \label{eq:NS} \eeq where $p$ is the pressure,
$\vec f$ is the external force, and $\nu$ is the viscosity. That
is supplemented by the incompressibility condition \[ \dive \vec
u = 0\,, \] and in our case by periodic boundary conditions for
a finite spatial region $V$. Similar equations arise for the
vorticity $\vec\omega\equiv \curl\vec u$, by taking the curl of
(\ref{eq:NS}) \beq \frac{\ddd \vec \omega}{\ddd t} + (\vec u \cdot
\grad) \vec \omega - (\vec \omega \cdot  \grad) \vec u =
\nu\triangle\vec \omega + \vec g \label{eq:NSv} \eeq with $\vec
g=\curl \vec f$.
The energy of the system is given by the total kinetic energy \[
E = \int_V{\frac{u^2}{2} } \] while the enstrophy is defined as
\[\Omega = \int_V{\frac{\omega^2}{2} }\,. \] Its role will
become clearer later on.

Consider Cartesian coordinates $r = (x_1,x_2,x_3) = (x,y,z)$; the
two-dimensional case is conveniently represented by setting the
third component of the velocity equal to zero: $\vec u = (u_1,
u_2, 0)$. Therefore $\vec\omega = (0,0,\omega_3)$ is better
represented by a pseudo-scalar $\omega = \omega_3$.
Eq.~(\ref{eq:NSv}) thus acts on a single component and since $\vec
u$ and $\vec \omega$ are now perpendicular to each other, it is
further simplified by the vanishing of the term
$(\vec\omega\cdot\grad)\vec u$,
 \beq \frac{\ddd \omega}{\ddd t} +
(\vec u \cdot  \grad) \omega = \nu\triangle \omega + g
\label{eq:NS2dv} \eeq where $g = \ddd f_1/\ddd y - \ddd f_2 / \ddd
x$.  The pressure has disappeared but equation
\eqref{eq:NS2dv} is still nonlocal because $\vec u = {\cal
K}\omega$ for some Biot-Savart kernel $\cal K$.

We take our system bounded in a rectangular domain, where it is
useful to consider the Fourier transform \[
 \omega_k
= \frac 1{2\pi} \int_V e^{ik\cdot r} \omega(r)
\]
 These modes
satisfy  $\overline\omega_k = \omega_{-k}$ which will always be
understood.

Upon Fourier-transforming~(\ref{eq:NS2dv}) we thus get, for $k\in
\bbZ^2\setminus \{0\}$,
 \beq\label{nspre}
\frac{\ddd \omega_k}{\ddd t} - F_k(\omega) = - \nu \,k^2 \omega_k
+ g_k\, \eeq
 where
\begin{subequations}
 \beq
\label{eq:Fk1}
F_k(\omega) = \sum_{j + \ell = k} \Phi_{j,\ell} \,\omega_j\omega_\ell
 \eeq
with coefficients given by \[ \Phi_{j,\ell} \equiv
\frac{j_2\ell_1 - j_1\ell_2}{4\pi}\,\big(\frac 1{|j|^2} -
\frac 1{|\ell|^2}\big)\;. \]

 These $\Phi_{j,\ell}= 0$ if and only if
either $j\parallel\ell$ or $|j|=|\ell|$.
Alternatively,
 \beq
\label{eq:Fk2}
F_k(\omega) =\sum_{0\neq \ell\neq k} \phi_{k,\ell}
\;\omega_\ell \omega_{k-\ell}  \eeq
with
\[
\phi_{k,\ell} \equiv
 \frac{k_1 \ell_2 - \ell_1 k_2}{4\pi} \left(\frac{1}{|\ell|^2}
- \frac{1}{|k-\ell|^2}\right) = \phi_{-k,-\ell}
\]
In that notation, \[ \phi_{k,\ell} + \phi_{k,k-\ell} = 0
\] represents the so called triad relation of~\cite{kraichnan75}.
Another alternative, which will be useful later, is
\beq \label{eq:Fk3}
F_k(\omega) = -\frac{1}{2\pi} \sum_{\ell} \frac{k_1 \ell_2 - \ell_1
k_2}{|k-\ell|^2} \omega_{\ell} \omega_{k-\ell} \;.
\eeq
\end{subequations}

Finally we must specify the forcing $g_k$. A translationally
invariant and stationary turbulent state can be achieved by
imposing a force that is homogeneous in space and time. A Gaussian
random field with zero mean is the simplest example: in that case
the force $\vec{f}$ in \eqref{eq:NS} is a Gaussian noise that is
white in time and colored in space, completely determined by its
covariance \[ \ave{ f_i(s,r) f_j(t,r') }=
C_{ij}(r-r')\delta(t-s)\,, \]
 where $\ddd_i C_{ij} =0$
(incompressibility).
Equation \eqref{nspre} then turns into
the stochastically driven NS \beq d\omega_k(t) = -\nu k^2 \omega_k
dt + F_k(\omega) dt + \sqrt{2\gamma_k} dW_k(t) \label{eq:omega}
\eeq in which $dW_k= d\overline W_{-k}$ represents a standard Wiener
process~\cite{LipsterShiryayev_book}.
That driving pumps vorticity into the system at
wavenumber $k$ with intensity $\gamma_k\geq 0$ while the viscosity
$\nu>0$ enters in the first term on the right-hand side of
\eqref{eq:omega} to dissipate the vorticity. Equation
\eqref{eq:omega} is the starting point of our analysis.

\subsection{Mathematical assumptions} The point of departure
\eqref{eq:omega} is a stochastic differential equation to be
understood in the It\^o-sense~\cite{LipsterShiryayev_book}.
Solutions are Markov processes but
note that they are infinite-dimensional.  In general the resulting
diffusion is not elliptic because some $\gamma_k$ can be made
zero.  That brings us to the problem of understanding the
assumptions on the strengths $\gamma_k$ and on the viscosity $\nu$
so that there is a unique invariant probability measure $\mu$.  A
lot of mathematical work has been devoted to that problem in
recent years.  For example~\cite{Mattingly1999}, if $\nu$ is
sufficiently large as a function of $\sum \gamma_k^2$, then $\mu$
is unique. Also~\cite{EckmannHairer2000}, $\mu$ is unique when
there is $\kappa>0$ so that $\gamma_k \sim |k|^{-\alpha}$ for some
$\alpha$ and for every $|k|>\kappa$.
Even~\cite{MattinglySinai2001,Kuksin2002,BricmontKupiainenLefevere2002},
when $\gamma_k\neq 0$ for every $|k|\leq N$ where $N$ is some
number that depends on $\nu$ and on $\sum \gamma_k^2$, then $\mu$
is unique.  We refer to \cite{HairerMattingly2004} for even
stronger and more recent results.

In what follows we simply assume, with no further ado, that $\mu$
is unique and has smooth local densities.
Another assumption in the technical manipulations is to start from
a finite dimensional analysis.  In other words, we choose a finite
but arbitrary $N$ and consider equation \eqref{eq:omega} only
for $k^2\leq N$ with $\gamma_k\neq 0$ there.  That cut-off will
take care of convergence problems in what follows and it allows us
to speak of $\mu(\omega)$ as the density of $\mu$ with respect to
the flat measure $d\omega$.

\subsection{Euler equation}
The vorticity and the corresponding enstrophy play an important
role in two-dimensional turbulence because of the appearance of an
extra conservation law.  The Euler equation corresponding to
\eqref{eq:omega} is \[ d\omega_k(t) = F_k(\omega) \,dt \;.\]
cf. \eqref{eq:Fk1}--\eqref{eq:Fk3}.
 It is easy to see that
 \beq\label{217}
  \sum_k
\overline\omega_k F_k(\omega) = 0 \eeq so that enstrophy is conserved
\[ \frac{d\Omega}{dt} = \sum_k \frac{1}{2}
\frac{d|\omega_k|^2}{dt}  = 0\;.\]
 As a consequence, the enstrophy
in \eqref{eq:omega} is changed by the injection (at rate
$\gamma_k$) and the dissipation (with intensity $\nu$) but is
transported without dissipation over the various modes via the
nonlinear and highly nonlocal terms in $F_k$.  That invites the
definition of various enstrophy currents.

\subsection{Currents}
The net enstrophy current $J_k$ that leaves the system at
wavenumber $k$ is obtained from investigating the sources and
sinks to the enstrophy. The total enstrophy dissipation over the
time interval $[-\tau,\tau]$ is computed from
\[
\Omega(t) = \frac 1{2} \sum_k |\omega_k|^2(t)
\]
and
 \[
\begin{split}
 \Omega(\tau) - \Omega(-\tau) =&
\sum_k
\bigg[ -\nu k^2 \int_{-\tau}^{\tau}|\omega_k|^2(t) dt + \\
& + \sqrt{2 \gamma_k}\;  \Re \int_{-\tau}^{\tau}
\overline\omega_k(t)\circ dW_k(t)
\bigg]
\end{split}
\]
 where the last integral is in the Stratonovich
 sense~\cite{LipsterShiryayev_book},
and $\Re$ stands for real part. The expression thus evaluates the
change of enstrophy during a time interval $[-\tau,\tau]$ for a
history $(\omega_k(t))$ and for a realization of the noise
$(dW_k(t))$.  It is therefore natural to put
\begin{eqnarray*}
J_k^{\text
{out}}&=& \nu k^2\int_{-\tau}^\tau |\omega_k|^2 dt
\\
J_k^{\text {in}}&=& \sqrt{2\gamma_k}\;
\Re\int_{-\tau}^\tau\overline\omega_k \circ dW_k
\end{eqnarray*}
as the current going ``out'', respectively ``in'' the system at
mode $k$ with respect to the external enstrophy reservoir.  The difference,
\[J_k \equiv J_k^{\text {out}} - J_k^{\text {in}}\]
is the net enstrophy current that leaves the system
(enters the environment) at mode $k$.

On the other hand the local conservation law reads
\begin{eqnarray}
\frac{|\omega_k|^2(\tau)}{2} - \frac{|\omega_k|^2(-\tau)}{2}
&=& -\nu k^2 \int_{-\tau}^{\tau}|\omega_k|^2(t)dt  \nonumber\\
&& + \Re\int_{-\tau}^{\tau}\overline\omega_k F_k(\omega) dt  \nonumber\\
&&+ \sqrt{2\gamma_k}\Re\int_{-\tau}^{\tau}\overline\omega_k\circ dW_k(t)
\nonumber\\
&=& -J_k^{\text {out}} + \sum_\ell J_{\ell k} + J_k^{\text {in}}
\label{221}
\end{eqnarray}
which defines
\[
J_{\ell k}\equiv  \frac{1}{2\pi} \Re
\int_{-\tau}^\tau dt\; \frac{\ell_1 k_2 - k_1 \ell_2 }{|\ell-k|^2}
\,\omega_{-k} \omega_{k-\ell} \omega_{\ell}
 \]
 the net current from mode $\ell$ to mode $k$ [here we used (\ref{eq:Fk3})].
 Note the asymmetry $J_{k\ell} = - J_{\ell k}$.

 As said before,  the
redistribution of enstrophy due to interactions between different
modes globally does not change the total amount of enstrophy in
the system: $\sum_k\sum_\ell J_{k\ell} =
0$, see \eqref{217}.

One of the main problems for the cascade picture is to understand
the direction of the flow of the $J_{k\ell}$.
That is basically determined by the stationary $\langle J_k\rangle$.
At the end of the paper we also discuss its fluctuations.

\subsection{Spectral distribution}\label{specdis}
Heuristically, the reason why the enstrophy flows towards small
scales (large wavenumber $k$) is because at these small scales the
dissipative term $\nu\triangle \omega$ in \eqref{eq:NS2dv}
dominates over the advection term $(\vec u \cdot \grad) \omega$. A
more refined argument, started by Kraichnan~\cite{kraichnan67:_t2d},
derives the two-dimensional cascade picture (the so called direct
cascade for the enstrophy and the inverse cascade for the energy)
by investigating the energy spectra.
The Fourier spectrum of energy embodies the KLB picture by showing a
power-law regime for each of the two cascades.
Since the enstrophy spectrum is simply related to the energy one,
from the inspection of the energy spectrum one can argue where energy and
enstrophy are transferred or dissipated. In a way the cascade of
enstrophy to small scales is the two-dimensional analogue of the
energy cascade in three dimensions. We skip here the details of
that Kolmogorov-Kraichnan theory as they have been excellently
reviewed by many.  We refer to~\cite{Kupiainen:_lecture,Bernard00:_amateurs}.

 The purpose of the
present paper is to try an alternative to that analysis by opening
an analogy with heat conduction.  It is interesting that in this
way a natural enstrophy dissipation function appears, the
thermodynamic entropy production, as will be explained in
Section~\ref{sec:dissip}.

\section{Formal analogy with heat dissipation}
\label{sec:analogy}
Remember our starting equation (\ref{eq:omega}).  Let us first
forget about the coupling between the various modes so that the
system is reduced to  the stochastic dynamics \beq d\omega_k(t) =
- \nu k^2 \omega_k(t) dt + \sqrt{2 \gamma_k} dW_k(t)
\label{eq:omega0} \eeq describing an ensemble of uncoupled
oscillators labeled by the wavenumber $k$. While in the original
NS equation the viscosity represents an irreversible loss, here it
balances reversibly with the stochastic forcing.  The dynamics
(\ref{eq:omega0}) has a reversible equilibrium measure \beq
\label{rev} \mu^0(d\omega) = \prod_k \frac{e^{-|\omega_k|^2 / 2
T_k}}{Z_k}\, d\omega_k d\overline\omega_k
\eeq that we will use as a reference.

 The parameter
$T_k \equiv \gamma_k /(\nu k^2)$ can be viewed as  a kind of
``temperature''  of the reservoir attached to wavenumber $k$; it
is of course no physical temperature. Thus, our approach is different from
previous attempts to use a thermodynamical formalism in turbulence,
identifying variables like $\omega_k^2$ with a temperature
(see~\cite{crisanti} and reference therein).

The reversibility of the
dynamics (\ref{eq:omega0}) is taken with the usual kinematical
time-reversal that reverses the sign of the velocity field: the
dynamical time-reversal of a history
\beq \label{33} \xi =
(\omega(t), t\in [-\tau,\tau]) \eeq in a given time interval
$[-\tau,\tau]$ is
\beq \label{34} \Theta\xi = (-\omega(-t), t\in
[-\tau,\tau]) \eeq

When we add the $F_k(\omega)$ to (\ref{eq:omega0}) to obtain
(\ref{eq:omega}) the oscillators become coupled, in fact in a
nonlinear and nonlocal way.  That coupling does however preserve
the enstrophy very much like a Hamiltonian coupling that conserves
the energy.  The picture that thus emerges is formally equivalent
to a heat conduction network where the vertices of the network are
represented by the modes $k$.

The Euler equation represents the conservative part of the
time-evolution.  That is changed by the addition of the Langevin
forces that represent ``thermal'' reservoirs at each of the $k$;
thus obtaining our equation (\ref{eq:omega}). Observe that the
``friction'' depends on the ``location'' $k$ of the oscillator.
Standard thermodynamics then teaches us that there will be a
``heat current'' from higher to lower temperature. That ``heat
current'' is in our present set-up played by the enstrophy
current.  Hence, if the driving makes $T_k$ a decreasing function
of $|k|$, e.g. by having $\gamma_k \sim k^{-\alpha}$ for some
$\alpha>0$, then, the enstrophy should be transported from small
$|k|$ towards larger $|k|$.  That picture will be detailed in the
following sections.

The forward generator $\mathcal{L}^+$ corresponding to the Markov
diffusion \eqref{eq:omega} can be split into a ``conservative''
and a ``dissipative'' part, $\mathcal{L}^+ = \mathcal{L}^+_c +
\mathcal{L}^+_d$, with
\begin{subequations}
\begin{align}\label{eq: hamilt-gen}
  \mathcal{L}^+_c \rho &= -\sum_k F_k(\omega)\,\frac{\partial \rho}{\partial \omega_k}
\\ \intertext{and}
  \mathcal{L}^+_d \rho &= \nu\sum_{k} k^2  \frac{\partial}{\partial \omega_k} (\omega_k \rho)
    + \sum_k \gamma_k \frac{\partial^2 \rho}{\partial \omega_k^2}
  = \sum_{k} \gamma_k \frac{\partial X_k}{\partial \omega_k}
\end{align}
\end{subequations}
where we made use of the shorthand \[ X_k \equiv e^{-\beta_k
|\omega_k|^2 / 2} \frac{\partial}{\partial \omega_k} (e^{\beta_k
|\omega_k|^2 / 2} \rho)\]
with $\beta_k \equiv \nu k^2/\gamma_k$.

For the stationary measure $\mu$ we have $\mathcal{L}^+\mu=0$ and
in particular
\beq\label{detbal}
 \langle F_k(\omega)
\overline\omega_k\rangle - \nu k^2 \langle |\omega_k|^2\rangle +
\gamma_k = 0 \;. \eeq
From now on we use that notation
$\langle\cdot\rangle$ to denote a stationary average according to
$\mu$. Equation \eqref{detbal} gives, for every time interval
$[-\tau,\tau]$,
 \beq\label{38}
\frac 1{2\tau}\,\ave{J_k} = \nu k^2(\ave{|\omega_k|^2}-T_k)\eeq
That equation is the detailed enstrophy balance equation in
stationarity; summing over $k$ gives the somewhat more familiar
\[
\nu\, \sum_k k^2\, \langle |\omega_k|^2\rangle =\sum_k  \gamma_k
\]
but at the same time and as a new interpretation of \eqref{38} we
recognize how the net current into the enstrophy reservoir at mode
$k$ is like a heat current into a thermal reservoir as determined
by the difference between, what now plays the role of a local
kinetic temperature, $\ave{|\omega_k|^2}$ and the reservoir
temperature $T_k$.

\section{Enstrophy dissipation}
\label{sec:dissip}
Continuing with the analogy above a quantity is now brought to the
forefront which we call the entropy current $S$. Since the net
enstrophy current leaving the system at each mode $k$ is $J_k=
J_k^{\text {out}} - J_k^{\text {in}}$ and the corresponding
``effective temperature'' is $T_k$ we put \beq\label{entropy}
 S \equiv
\sum_k \frac 1{T_k} J_k
\eeq
 as variable entropy current.  It is a function of the history \eqref{33}
 over $[-\tau,\tau]$.
 The entropy current $S$ is the entropy production in the environment
 associated to the enstrophy dissipation; it is the usual sum over all
 dissipative currents divided by the respective temperatures.  In
 the stationary state, the average $\langle S \rangle$ is the
 total change of the entropy in the universe over the time-interval
 $[-\tau,\tau]$.

   We will now show what is suggested
 thermodynamically by the previous analogy: $S$ should measure the irreversibility and
 $\langle S \rangle \geq 0$ as a consequence of the Second Law of
Thermodynamics.

Remember that $\mu$ is the
 stationary measure of the
NS dynamics~(\ref{eq:omega}); we denote by $\pi \mu$ its
time-reversal. In a given time interval $[-\tau,\tau]$
each history \eqref{33} is
realized in the system with a probability that comes from the
path-space measure $P_\mu^\tau(d\xi)$, i.e., the stationary Markov
diffusion process associated to the stationary measure $\mu$ and
the stochastic
dynamics \eqref{eq:omega}.

We compute the logarithmic density  (see also
\eqref{33}-\eqref{34})
 \beq
\label{rrr}
R\equiv \ln \frac{P_\mu^\tau}{P_{\pi\mu}^\tau\Theta}
\eeq
as a measure of
irreversibility.  It gives the ratio between the probability of a
history $\xi$ and the probability of the time-reversed history
$\Theta\xi$.  We show that $R$ coincides with $S$
up to a temporal boundary term.  Moreover, taking  stationary
averages $\langle R \rangle = \langle S \rangle$. Since by
construction, $\langle R
\rangle \geq 0$ it also follows that $\langle S \rangle \geq 0$.

To compute $R$ it is useful to compare the path-space measure with
the reference path-space measure of the uncoupled case
(\ref{eq:omega0}), denoted by $P_{\mu^0}^{0,\tau}$ (that one is
stationary and reversible). Thus first we compute the action
\[
A_\mu \equiv \ln \frac{P_\mu^\tau}{P_{\mu^0}^{0,\tau}} \]
and
similarly $A_{\pi\mu}\circ\Theta$, to finally estimate \eqref{rrr}
as the source of time-reversal breaking
\beq R=A_\mu - A_{\pi\mu}\, { \circ\, \Theta } \label{eq:R_via_A} \eeq

 The comparison of the
two measures $P$ and $P^0$ is made by means of the Girsanov
formula~\cite{LipsterShiryayev_book}, obtaining
\begin{widetext}
\begin{eqnarray*}
A_\mu =&& \sum_k \frac{1}{2\gamma_k} \left\{ \int_{-\tau}^\tau
\left[ \nu k^2 \Re(\omega_k \overline F_k(\omega)) -
\frac{1}{2}|F_k(\omega)|^2 \right] dt +\Re\left[ \int_{-\tau}^\tau
\overline F_k(\omega) d\omega_k\right] \right\} + \ln
\mu(\omega(-\tau)) - \ln \mu^0(\omega(-\tau)) \;.
\end{eqnarray*}
Substituting $\Theta\xi$ gives
\begin{eqnarray*}
A_{\pi\mu}\circ\Theta = && \sum_k \frac{1}{2\gamma_k} \left\{
\int_{-\tau}^\tau \left[ -\nu k^2 \Re(\omega_k \overline F_k(\omega)) -
\frac{1}{2}|F_k(\omega)|^2 \right] dt +\Re\left[ \int_{-\tau}^\tau
\overline F_k(\omega) d\omega_k\right]\circ\Theta \right\}
 + \ln \mu(\omega(\tau)) - \ln \mu^0(\omega(\tau))\,.
\end{eqnarray*}
Here It\^o stochastic integrals are performed and one should
remember that these are themselves not time-reversal
symmetric~\cite{LipsterShiryayev_book}.
As an example for computing \eqref{eq:R_via_A} we see that
\begin{eqnarray*}
\Re\left[ \int_{-\tau}^\tau \overline F_k(\omega) d\omega_k \right] -
\Re\left[ \int_{-\tau}^\tau \overline F_k(\omega) d\omega_k \right]
\circ\Theta &=&\\
\lim_{\Delta t\to 0} \Re
\left[
\sum_j \overline
F_k(\omega(t_{j-1}))[\omega_k(t_j) - \omega_k(t_{j-1})]
\right]
- \lim_{\Delta t\to 0} \Re
\left[
\sum_j \overline F_k(-\omega(t_{j}))[
- \omega_k(t_{j-1}) + \omega_k(t_j)]
\right] &=&\\
- \lim_{\Delta t\to 0} \Re
\left[\sum_j 
\frac{\overline F_k(\omega(t_{j}))- \overline F_k(\omega(t_{j-1}))}
{\overline\omega_k(t_j) - \overline\omega_k(t_{j-1})}
[\omega_k(t_j) - \omega_k(t_{j-1})]^2
\right] &=&\\
-\Re\left[ 
\int_{-\tau}^\tau \frac{\partial \overline F_k(\omega)}{\partial\overline\omega_k} dt
\right]&=& 0 \end{eqnarray*}
because $\partial\overline F_k / \partial \overline\omega_k =
0$, see \eqref{eq:Fk2}.
\end{widetext}

As a consequence, \eqref{eq:R_via_A} becomes \beq R = S + \ln
\mu(\omega(-\tau)) - \ln \mu(\omega(\tau)) \label{eq:R} \eeq with
\[ S = \sum_k \frac{1}{T_k}\left\{
\left[\frac{\omega_k^2(-\tau)}{2} -\frac{\omega_k^2(\tau)}{2}\right]
+\int_{-\tau}^\tau\Re[\omega_k \overline F_k(\omega)] dt \right\}\]
From \eqref{221} that
expression coincides exactly with \eqref{entropy}, as promised.

 When instead of the stationary $\mu$ we had
taken  some initial density evolving as $\rho_t, t\in
[-\tau,\tau]$, the analysis above would be essentially unchanged.
 In that case the source of irreversibility is
\begin{equation}\label{entppt}
R = \sum_{k} \frac 1{T_k} J_k + \ln\rho_{-\tau}(\omega_{-\tau})
-\ln\rho_{\tau}(\omega_\tau)
\end{equation}
where the only difference with \eqref{eq:R} resides in the last
two terms, the temporal boundary
\[
[-\ln\rho_\tau(\omega_{\tau})] -
[-\ln\rho_{-\tau}(\omega_{-\tau})]
\]

\subsection{Mean entropy production}
From the definition (\ref{rrr}) it directly follows
\[
 \langle e^{-R}\rangle = 1
\]
(it is essentially the normalization condition of the path-space measure
$P_{\pi\mu}^\tau \Theta$).
 Hence, by a convexity
inequality, the stationary enstrophy dissipation $\langle S
\rangle = \langle R \rangle \geq 0$. We can however be more
explicit concerning that point by deriving an expression for
$\langle S \rangle$ which is explicitly non-negative.  In fact, we
will show that \beq \label{strict} \langle S \rangle = \sum_k
\gamma_k \langle \Big(\exp[-V_k(\omega)]\frac{\partial}{\partial
\omega_k} \exp[V_k(\omega)]\Big)^2 \rangle\eeq where
 \[ V_k(\omega)\equiv
|\omega_k|^2/(2T_k) + \ln \mu(\omega)
\]
From \eqref{strict}, $\langle S \rangle > 0$ strictly as we can
only have
\[
\frac{\partial}{\partial \omega_k}\left[
e^{\beta_k|\omega_k|^2/2} \mu(\omega)
\right] = 0
\]
for all $k$ when $\mu=\mu_0$ of \eqref{rev}.

Here comes the proof of \eqref{strict}. Denote by
$\bbE_{\rho_{-\tau}}$ the expectation in the process
$P_{\rho_{-\tau}}^\tau$ started from $\rho_{-\tau}$. We assume
that at time $\tau$ the evolved measure is described by a density
$\rho_\tau$.  We have \eqref{entppt}, in expectation,
\begin{equation}\label{inexp}
  \bbE_{\rho_{-\tau}} [R]  =
  \sum_{k} \beta_k \bbE_{\rho_{-\tau}}[ J_k ]
     + S(\rho_\tau) - S(\rho_{-\tau})
\end{equation}
where $S(\rho) \equiv - \int\, d\omega \, \rho(\omega) \ln
\rho(\omega)$ is the Shannon entropy of the density $\rho$.
Another formulation is
\[
\bbE_{\rho_{-\tau}} [R] = \int_{-\tau}^{\tau} \dot R(t)\, dt
\]
with, similar to \eqref{38},
\begin{equation}\label{trans}
\dot R(t) \equiv \nu \sum_{k} k^2\beta_k [\int d\omega\,
|\omega_k|^2 \rho_t(\omega) - \frac 1{\beta_k}]
     + \frac{d}{dt}S(\rho_t)
\end{equation}
The previous considerations thus identify the mean  dissipation
rate at time $t$ (in the transient regime)
with $\dot R(t)$.

To see the relation with \eqref{strict} we start by evaluating the
time-derivative of the Shannon entropy:
\begin{equation}
  \frac{dS}{dt}(\rho) = -\int d\omega\, \frac{d\rho}{dt} \ln\rho
  = -\int d\omega\, (\mathcal{L}^+ \rho) \ln\rho
\end{equation}
Using the invariance of the Shannon entropy under the conservative
(Euler) part of \eqref{eq: hamilt-gen}, we get
\begin{equation}\label{eq: Shannon}
\begin{split}
  \frac{dS}{dt}(\rho) &= -\int d\omega\, (\mathcal{L}^+_d \rho) \ln\rho
  = \sum_{k} \gamma_k
    \int d\omega\, X_k \frac{\partial \ln \rho}{\partial \omega_k}
\\
  &= \sum_{k} \gamma_k
    \int d\omega\, X_k \Bigl( \frac{X_k}{\rho} - \beta_k \overline\omega_k  \Bigr)
    \\
    &= \sum_k \gamma_k \langle[\frac{X_k}{\rho}]^2\rangle -
    \nu\sum_k k^2\int d\omega\,\overline \omega_k X_k
\end{split}
\end{equation}
Minus the second term reads
\begin{equation} \label{minus2}
\begin{split}
  \nu \sum_{k} k^2&\int d\omega\, \overline \omega_k X_k
  = \nu \sum_{k} k^2 \int d\omega\, \overline \omega_k
    \Bigl( \frac{\partial \rho}{\partial \omega_k} + \beta_k \omega_k \rho \Bigr)
\\
  &= \nu \sum_{k} \beta_k k^2 \int d\omega\, \rho
     \Bigl( |\omega_k|^2 - \frac{1}{\beta_k} \Bigr)
\end{split}
\end{equation}
Substituting \eqref{minus2} into \eqref{eq: Shannon} and then
\eqref{eq: Shannon} into \eqref{trans}, we immediately obtain the
desired identity \eqref{strict}.

\subsection{Enstrophy network} The situation can now be summarized as
follows: locally, in the stationary measure, we have
 \[
\sum_\ell \ave{J_{\ell k}} =  \ave{J_k} =
\ave{\overline \omega_k \,F_k(\omega)} \]
 and globally \beq\label{golbal}  \sum_k \ave{J_k} = 0
 \eeq
 For the enstrophy dissipation
\beq
\label{ed} \langle S\rangle = \sum_k \beta_k
\ave{J_k} > 0\;.
\eeq
We have here formally the same situation as for a heat
conduction network as considered e.g.\ in~\cite{heat-conduction-net}.
The relations
\eqref{golbal} and \eqref{ed} do not of course uniquely determine
the mean  enstrophy currents but their direction or sign is thermodynamically
 determined by analogy with heat conduction.

 Let us first consider
the typical case where the strengths $\gamma_k$ are non-zero only
for a neighborhood of $k=0$, say $\gamma_k=1$ when $|k|\leq \eta$
and outside that large wavelength regime, $\gamma_k\downarrow 0,
|k|>\eta$. In terms of heat conduction it would mean that the
temperatures $T_k = 1/(\nu k^2)$ are decreasing outward in the
disk for $|k|\leq \eta$ and fall to $T_k=0$ outside ($|k|>\eta$).
Clearly then, there will be a heat current toward increasing
$|k|$ or, here, an enstrophy current towards smaller wavelengths.
In other words, the enstrophy current is a kind of nonlocal heat
current the direction of which is determined by the Second Law.
Because of the nonlocality of the term $F_k(\omega)$ the current
will not stop at the boundary of the disk but will be more and
more suppressed when regarding $J_{k\ell}$ for $k$ inside and
$\ell$ outside the disk. For really large $\ell$ there is no
longer a visible local heat current. That seems compatible with the
observations~\cite{tran3} that the enstrophy cascade
remains pretty localized around the forcing window.

In general however, when all $\gamma_k>0$ are active, we have  a
``temperature'' profile $\gamma_k/(\nu k^2)$ that can of course be
complicated.  If the $\gamma_k$ only depend on $|k|$ we have in
essence a one-dimensional heat conduction problem (along the
radial direction).

\subsection{Fluctuations}

Looking back at \eqref{inexp} and \eqref{trans}, we found the mean
entropy as the change of entropy in the environment $S$ plus the
change of (Shannon) entropy due to the stochastic dynamics in the
system.  Its stationary mean $\langle R \rangle > 0$ is strictly
positive.  We will now look at its fluctuations.  More precisely,
we consider the $R$ of \eqref{eq:R} and ask for its probability
distribution.  Since by construction \eqref{rrr},
\[
P_\mu^\tau(\xi) = e^{R(\xi)}\;P_{\pi\mu}^\tau(\Theta\xi)
\]
we have that
\beq\label{fs}
 \int\, f(\Theta\xi)\,dP_{\pi\mu}^\tau
= \int\, f(\xi)\,e^{-R(\xi)}\,dP_\mu^\tau(\xi)
\end{equation}
is exactly valid for all observations $f$ and for all times
$\tau$.  The relation \eqref{fs} is called a fluctuation symmetry,
see e.g.~\cite{Maes04:_poincare}, because it generates the so called
fluctuation theorem for the entropy production as first formulated
in~\cite{Evans93,GallavottiCohen95,GallavottiCohen95_JSP}.
Remember that $R$ equals the $S$ up to a
temporal boundary term, see \eqref{eq:R} and \eqref{entppt}.

One of the consequences of the fluctuation symmetry \eqref{fs} is
that
\beq\label{dinv}
 \frac{\mbox{Prob}[R < 0]}{\mbox{Prob}[R >
0]} = \langle e^{-R}|R>0\rangle
\eeq
 which is sometimes easier to
check numerically and experimentally. Roughly speaking, that last
relation tells us that the probability of observing the inverse
cascade for the enstrophy is exponentially smaller than the
probability of observing the direct cascade.

\section{Conclusions}
\label{sec:concl}

The main conclusion is derived from the analogy with a heat
conduction network.  Above and beyond all detailed physical
mechanisms that give rise to the direct enstrophy cascade in
two-dimensional turbulence stands the Second Law of Thermodynamics
for the entropy \eqref{entropy} which gives a direction to the
enstrophy flow.  The relevant parameter is the ratio
$\gamma_k/(\nu k^2)$ which plays the role of an effective
temperature of an enstrophy reservoir to which each mode $k$ is
coupled.  If the forcing is restricted to a finite window, then
the temperature outside is effectively equal to zero.  The
conservative part in the enstrophy conduction is nonlocal but does
not contribute to the dissipation.

We have identified a general entropy function \eqref{entropy} and
\eqref{eq:R},  also in the transient regime, see \eqref{inexp} and
\eqref{trans}.  We have shown that the stationary entropy
production is strictly positive.  It provides the general
mechanism driving the direct enstrophy cascade.  The fluctuations
in the entropy satisfy the symmetry \eqref{fs} which gives an
estimate \eqref{dinv} of the relative probabilities of direct
versus inverse cascades.

An important open question remains however.  The above analogy is
silent about the inverse energy cascade.  We have not found a heat
conduction analogue which would reveal the inverse cascade for the
energy dissipation in two-dimensional turbulence. Of course, as
energy and enstrophy are entangled and spectrally related, the
direct enstrophy cascade has direct consequences in the form of
the inverse energy cascade.  That point follows from the standard
treatments, see also Section~\ref{specdis}, 
as in~\cite{kraichnan67:_t2d} but has not been clarified in the present
paper. In fact, a naive extension of the present formalism but for
the energy would find the inverse cascade quite surprising as it
seems to reduce entropy. We have not investigated whether the
combination of dissipative currents, enstrophy and energy, would
still lead to a total positive entropy production, as expected
thermodynamically.  Clarifying that entropy balance remains one of
the most intriguing problems of two-dimensional turbulence.

\end{document}